\documentclass{article}

\usepackage{authblk}
\usepackage{amsmath}
\usepackage{dsfont} 
\usepackage{amssymb}
\usepackage{esvect} 
\usepackage[colorlinks=true,linkbordercolor={1 0 0}]{hyperref} 

\newcommand{\hf}{\mathcal{H}^{(F)}}
\newcommand{\hs}{\mathcal{H}^{(S)}}

\renewcommand{\k}{{\vv{k}}}

\title{Crypto-Hermitian Approach to the Klein-Gordon Equation}
\date{}
\author[1,2]{Iveta Semor\'adov\'a}
\affil[1]{Nuclear Physics Institute, Czech Academy of Science, \v{R}e\v{z} near Prague, Czech Republic}
\affil[2]{Faculty of Nuclear Science and Physical Engineering, Czech Technical University in Prague, Czech Republic}

\begin{document}

\maketitle
\vspace{-2em}
{E-mail: semorive@fjfi.cvut.cz}

\begin{abstract}
We explore the Klein-Gordon equation in the framework of crypto-Hermitian quantum mechanics. Solutions to common problems with probability interpretation and indefinite inner product of the Klein-Gordon equation are proposed.
\end{abstract}

\section{Introduction}
The urge to unite special theory of relativity with quantum theory emerged shortly after their discovery. The first relativistic wave equation was introduced in 1926 simultaneously by Klein \cite{Klein}, Gordon \cite{Gordon1}, Kudar \cite{Kudar}, Fock \cite{Fock1}\cite{Fock2} and de Donder and Van Dungen \cite{Donder}. Schr\"{o}dinger himself formulated it earlier in his notes together with the Schr\"{o}dinger equation \cite{Schrodinger}. However, with the introduction of the Klein-Gordon equation arose several problems. For given momentum equation allows solutions with both positive and negative energy, it has an extra degree of freedom due to presence of both first and second derivatives and mainly its density function is indefinite and therefore cannot be consistently interpreted as probability density. Also, the
predictions based on this equation seemed to disagree with experiments (cf., e.g., the historical remark in \cite{constantinescu2013problems}). Therefore, few years later all the attention shifted to the Dirac equation.

More than ninety years old problem of proper probability interpretation of the Klein-Gordon equation was first solved in 1934 by Pauli and Weisskopf \cite{pauli1934} by reinterpreting the Klein-Gordon equation in the context of quantum field theory. Quantum mechanical approach to the Klein-Gordon equation was forgotten until Ali Mostafazadeh brought it back in 2003 \cite{MFKGtype}. In his work, he made use of pseudo or quasi-Hermitian approach to quantum mechanics.

Mathematical ideas of quasi-Hermitian theory originate from works of Dieudonn\'e \cite{die} and Dyson \cite{dyson}, though it wasn't until 1992 when the theory was consistently explained and applied in nuclear physics by Scholtz, Geyer and Hahne \cite{SGH}. This groundbreaking work initiated fast growth of interest popularized in 1998 by Bender and Boettcher \cite{BB}. Nowadays the application of the theory is moving away from quantum mechanics to other branches of physics, such as optics. 

We would like to return to the problem of proper interpretation of the Klein-Gordon equation in the framework of Quantum mechanics only. Several publications concerning this subject appeared \cite{MFKG, mostafazadeh2006quantum1, mostafazadeh2006quantum2, mostafazadeh2006physical} or \cite{znojil2004relativistic, MZdots, znojil2004pseudo}. But even these studies did not provide an ultimate answer to all
of the open questions. Some of them will be addressed in what follows. 

\section{Klein-Gordon equation in Schr\"odinger form}
The Klein-Gordon equation for free particle can be written in common form 
\begin{equation}
(\square+\frac{m^2c^2}{\hbar^2})\psi(t,x)=0\ , 
\label{KG}
\end{equation}
where $\square=\frac{1}{c^2}\partial_t^2-\Delta=\partial_{\mu}\partial^{\mu}$ is the d'Alembert operator.
From now on we will use the natural units $c=\hbar=1$, furthermore we can denote $K=-\Delta+m^2$ and rewrite (\ref{KG}) as
\begin{align}
	(i\partial_t)^2\psi(t,x)=K\psi(t,x)\ . \label{KG2}
\end{align}
The fact that the Klein-Gordon equation is differential equation of second order in time gives it an extra degree of freedom. Feshbach and Villars \cite{FV} suggested solution to this problem by introducing two-component wave function and therefore making the extra degree of freedom more visible. Following their ideas together with even earlier ideas of Foldy \cite{foldy1956synthesis}, we can replace the Klein-Gordon equation with two differential equation of first order in time.
Inspired by convention introduced in \cite{znojil2004relativistic} we put
\begin{equation}
\Psi^{(1)}=i\partial_t\psi\ ,\quad \Psi^{(2)}=\psi\ . 
\end{equation}
Now, equation \eqref{KG2} can be decomposed into a pair of partial differential equations  
\begin{align}
i\partial_t\Psi^{(1)}&=K\Psi^{(2)}\ ,\\
i\partial_t\Psi^{(2)}&=\Psi^{(1)}\ ,
\end{align}
which, written in the matrix form, become
\begin{align}
i\partial_t\begin{pmatrix}\Psi^{(1)}\\\Psi^{(2)}\end{pmatrix}=\begin{pmatrix}0&K\\I&0\end{pmatrix}\begin{pmatrix}\Psi^{(1)}\\\Psi^{(2)}\end{pmatrix}\ . \label{FV}
\end{align}
Hamiltonian of the quantum system takes form
\begin{equation}
H=\begin{pmatrix}0 & K \\ 1 & 0\end{pmatrix}\ ,
\label{KGhamiltonian}
\end{equation}
and enters the Schr\"{o}dinger equation
\begin{align}
	i\partial_t\Psi(t,x)=H\Psi(t,x)\ ,\quad \Psi=\begin{pmatrix}\Psi^{(1)}\\\Psi^{(2)}\end{pmatrix}\ .\label{KGSCH}
\end{align}
Two-component vectors $\Psi(t)$ belong to 
\begin{equation}
\mathcal{H}=L^2(\mathds{R}^3)\oplus L^2(\mathds{R}^3)
\end{equation}
and the Hamiltonian $H$ may be viewed as acting in $\mathcal{H}$. 

The so called Schr\"odinger form of the Klein-Gordon equation \eqref{KGSCH} is equivalent to the original Klein-Gordon equation \eqref{KG}. It is in more familiar form, although, new challenge arises with the manifest non-Hermiticity of Hamiltonian \eqref{KGhamiltonian}.

\subsection{Eigenvalues}

New form of the Klein-Gordon equation \eqref{KGSCH} has many benefits. One of them is simplification of calculation of its eigenvalues to mere solving the eigenvalue problem for operator $K$
\begin{equation}
K\psi_n=\epsilon_n\psi_n\ .
\label{Keigen}
\end{equation}
The relationship between eigenvalues $\epsilon_n$ of the operator $K$ and eigenvalues $E_n$ of the non-Hermitian operator $H$ of the Schr\"odinger form of the Klein-Gordon equation
\begin{align}	
	\begin{pmatrix}0&K\\I&0\end{pmatrix}\begin{pmatrix}\Psi^{(1)}\\\Psi^{(2)}\end{pmatrix}=E\begin{pmatrix}\Psi^{(1)}\\\Psi^{(2)}\end{pmatrix} 
	\label{eigen}
\end{align}
can be easily seen. Equation \eqref{eigen} is formed from two algebraic equations 
\begin{equation}
K\Psi^{(2)}=E\Psi^{(1)}\ ,\quad \Psi^{(1)}=E\Psi^{(2)}\ .
\end{equation}
After insertion of the second one to the first one we obtain
\begin{align}
	K\Psi^{(2)}_n=E_n^2\Psi^{(2)}_n\ ,
\end{align}
which compared with equation \eqref{Keigen} gives us following relation between eigenvalues
\begin{align}
	\epsilon_n=E_n^2\ .
	\label{eigen_relation}
\end{align}
We can see, that eigenvalues $E_n$ remain real under assumption of $\epsilon_n>0$.

Relationship between corresponding eigenvectors 
\begin{equation}
H\Psi^{(\pm)}_n=E_n^{(\pm)}\Psi^{(\pm)}_n\ ,\quad
\Psi^{(\pm)}_n=\begin{pmatrix}\pm\sqrt{\epsilon_n}\psi_n\\\psi_n\end{pmatrix}
\label{eigenvectorsH}
\end{equation}
is also easy to see.

\subsection{Free Klein-Gordon equation}

In case of free Klein-Gordon equation operator
\begin{equation}
	K=-\Delta+m^2
\end{equation}
acting on $\mathcal{H}=L^2(\mathds{R}^3)$ is positive and Hermitian. It has continuous and degenerate spectrum. As suggested in \cite{MFKGtype}, we identify the space $\mathds{R}^3$ with the volume of a cube of side $l$, as $l$ tends to infinity. Than we can treat the continuous spectrum of $K$ as the limit of the discrete spectrum corresponding to the approximation. The eigenvalues are given by
\begin{equation}
\epsilon_{\vv{k}}=k^2+m^2
\end{equation}
and corresponding eigenvectors $\psi_\k=\Psi^{(2)}_{\vv{k}}$ are
\begin{equation}
\psi_\k(\vv{x})=\langle\vv{x}|\k\rangle=(2\pi)^{-3/2}e^{i\vv{k}.\vv{x}}\ ,
\end{equation}
where $\k\in\mathds{R}^3$ and $\k.\k=k^2$. We can see that $\psi_\k\notin L^2(\mathds{R}^3)$. They are generalized eigenvectors, i.e. vectors which eventually becomes $0$ if $(K-\lambda I)$ is applied to it enough times successively, describing scattering states \cite{MFKGtype}.

Vectors $\psi_\k$ satisfy orthonormality and completeness conditions
\begin{equation}
\langle\k|\k'\rangle=\delta(\k-\k')\ ,\quad \int{d^3k|\k\rangle\langle\k|)}=1\ 
\end{equation}
and operator $K$ can be expressed by its spectral resolution as
\begin{equation}
K=\int{d^3k(k^2+m^2)|\k\rangle\langle\k|}\ .
\label{spectralK}
\end{equation}
From the relations \eqref{eigen_relation} and \eqref{eigenvectorsH} we see that eigenvalues and eigenvectors of $H$ are given by
\begin{equation}
E_{\vv{k}}^{(\pm)}=\pm\sqrt{\vv{k}^2+m^2}\ ,\quad 
\Psi_{\vv{k}}^{(\pm)}=\begin{pmatrix}\pm\sqrt{{\vv{k}}^2+m^2} \\ 1 \end{pmatrix}\psi_\k\ .
\end{equation}
The eigenvectors $\Phi^{(\pm)}_{\vv{k}}$ of adjoint operator $H^\dagger$ are
\begin{equation}
\Phi^{(\pm)}_{\vv{k}}=\begin{pmatrix}1\\ \pm\sqrt{{\vv{k}}^2+m^2}\end{pmatrix}\psi_{\vv{k}}\ ,
\label{eigenvectorHdagger}
\end{equation}
which form together with $\Psi_{\vv{k}}^{(\pm)}$ complete biorthogonal system 
\begin{equation}
\langle\Phi^{(\nu)}_{\vv{k}'}|\Psi_{\vv{k}}^{(\nu')}\rangle=\delta(\vv{k}-\vv{k}')\delta_{\nu\nu'}2E_\k^{(\nu)}\ ,
\end{equation}
where $\nu,\nu'=\pm1$.

\section{Crypto-Hermitian approach}

Apparent non-Hermiticity of Hamiltonian \eqref{KGhamiltonian} can be dealt with by means of the crypto-Hermitian theory (sometimes also called quasi-Hermitian \cite{MFshrnuti} or $\mathcal{PT}$-symmetric \cite{bender2007making}). 

Hamiltonian is non-Hermitian $H\neq H^\dagger$ only in the false Hilbert space $\hf=(V,\langle\cdot|\cdot\rangle)$. The underlying vector space of states is fixed, given by the physical system. However, we have a freedom in the choice of inner product. If we represent our Hamiltonian in different secondary Hilbert space $\hs=(V,\langle\langle\cdot|\cdot\rangle)$, with newly defined inner product
\begin{equation} 
\langle\langle\cdot|\cdot\rangle=\langle\varphi|\Theta|\psi\rangle\ ,
\label{inner}
\end{equation}
it may become Hermitian. So called metric operator $\Theta$ must be positive definite, everywhere-defined, Hermitian and bounded with bounded inverse. Operators for which such inner product exist will be called crypto-Hermitian (c.f. \cite{3HS}). They satisfy the so called Dieudon\'ee equation
\begin{equation}
H^\dagger\Theta=\Theta H
\label{Die}
\end{equation} 
and they are similar to Hermitian operators 
\begin{equation}
h=\Omega H\Omega^{-1}\ ,
\end{equation}
where $\Theta=\Omega^\dagger\Omega$ is invertible and $h=h^\dagger$.

In such scenario, the problem of negative probability interpretation of the Klein-Gordon equation can be reinterpreted as the problem of the wrong choice of metric operator $\Theta$. If we would be able to find more appropriate choice of representation space $\hs$, this problem would disappear.   

\subsection{Computation of the metric}

One of the possible ways how to construct metric operator $\Theta$ for given crypto-Hermitian Hamiltonian $H$ is by summing the spectral resolution series. It requires the solution of eigenvalue problem for $H^\dagger$. In what follows, we try to construct the metric operator for free Klein-Gordon equation 
\begin{equation}
\Theta=\int{d^3k\left(\alpha^{(+)}|\Phi^{(+)}_{\k}\rangle\langle\Phi^{(+)}_{\k}|+\alpha^{(-)}|\Phi^{(-)}_{\k}\rangle\langle\Phi^{(-)}_{\k}|\right)}\ ,
\end{equation}
where we insert eigenvectors $\Phi^{(\pm)}_\k$ as computed in \eqref{eigenvectorHdagger}
\begin{equation}
\Theta=\int{d^3k\begin{pmatrix}\alpha&\beta\sqrt{k^2+m^2}
\\\beta\sqrt{k^2+m^2}&\alpha(k^2+m^2)\end{pmatrix}|\k\rangle\langle\k|}\ ,
\end{equation}
where $\alpha=\alpha^{(+)}+\alpha^{(-)}$, $\beta=\alpha^{(+)}-\alpha^{(-)}$.
By means of equation \eqref{spectralK} we obtain family of metric operators
\begin{equation}
\Theta=\begin{pmatrix}\alpha&\beta K^{1/2}\\\beta K^{1/2}&\alpha K\end{pmatrix}\ ,
\label{metricKG}
\end{equation}
where 
\begin{equation}
K^{1/2}=\int{d^3k\sqrt{k^2+m^2}}|\k\rangle\langle\k|\ .
\end{equation}

With the knowledge of the metric operator (\ref{metricKG}), we can construct positive definite inner product defining Hilbert space $\mathcal{H}^{(S)}$
\begin{equation}
\begin{split}
\langle\langle\Psi|\Phi\rangle&=\alpha(\langle\psi|K|\varphi\rangle+\langle\dot{\psi}|\dot{\varphi}\rangle)\\
&+i\beta(\langle\psi|K^{1/2}|\dot{\varphi}\rangle-\langle\dot{\psi}|K^{1/2}|\varphi\rangle)\ ,
\end{split}
\label{innerproductKG}
\end{equation}
where $\dot{\varphi},\dot{\psi}$ denote corresponding time derivatives (In fact, this equation is just an
explicit version of equation (\ref{inner})).

\subsection{The discrete case}

Unfortunately, the metric operator (\ref{metricKG}) is unbounded and therefore doesn't satisfy all the requested properties we put upon metric operator. As was emphasized in \cite{kretschmer2004quasi}, boundedness of metric operator $\Theta$ is very important property, it guarantees that convergence of Cauchy sequences is not affected by introduction of new inner product (\ref{inner}). The possibility of the use of unbounded metrics is treated e.g. in the last chapter of \cite{ZnojilKniha}.

To overcome the problems with unboundedness of the metric operator \eqref{metricKG}, we choose to shift our attention to a discrete model.
In the discrete approximation the metric operator stays bounded.
We make use of equidistant, Runge-Kutta grid-point coordinates
\begin{equation}
x_k=kh\ ,\quad k=0,\pm1,\pm2\dots\ ,
\end{equation}
Laplacian can be expressed as
\begin{equation}
-\frac{\psi(x_{k+1})-2\psi(x_k)+\psi(x_{k-1})}{h^2}\ ,
\end{equation}
The explicit occurrence of the parameter $h$ will be important for the study of the continuum limit in which the value of $h$
would decrease to zero. Otherwise we may set $h=1$ in suitable units. Following further ideas from \cite{znojil2012n}, Laplace operator $\Delta$ can be discretized into matrix form
\begin{align}
\Delta^{(n)}=\begin{pmatrix}
2&-1&&&\\
-1&2&-1&&\\
&-1&2&\ddots&\\
& &\ddots&\ddots&-1 \\
&&&-1&2
\end{pmatrix}\ \label{laplace}
\end{align}
Matrix (\ref{laplace}) is Hermitian and therefore diagonalizable, i.e. similar to diagonal matrix. Hence for our purposes it is enough to compute with $n\times n$ real diagonal matrix
\begin{align}
K=
\begin{pmatrix}
a_1&0&\cdots&0 \\
0&a_2&\cdots&0 \\
\vdots&\vdots&\ddots&\vdots\\
0&0&\cdots&a_n
\end{pmatrix}\ .
\end{align} 

Let $A$, $B$, $C$ be real matrices $n\times n$, where $A=A^T$, $B=B^T$. Than we can write the Dieudonn\'{e} equation \eqref{Die} by means of block matrices
\begin{align}
\begin{pmatrix} 0&I\\K&0\end{pmatrix}
\begin{pmatrix}A&C^T\\C&B\end{pmatrix}
=
\begin{pmatrix}A&C^T\\C&B\end{pmatrix}
\begin{pmatrix}0&K\\I&0\end{pmatrix}\ .
\end{align} 

We obtain following conditions
\begin{align}
C=C^T,\ KC=C^TK,\ B=KA=AK\ .
\end{align}

Real symmetric matrix which commutes with diagonal matrix must be diagonal. Thus the form of our metric operator is as follows
\begin{align}
\Theta=\begin{pmatrix}
\alpha_1&\cdots&0&\beta_1&\cdots&0\\
\vdots&\ddots&\vdots&\vdots&\ddots&\vdots\\
0&\cdots&\alpha_n&0&\cdots&\beta_n\\
\beta_1&\cdots&0&a_1\alpha_1&\cdots&0\\
\vdots&\ddots&\vdots&\vdots&\ddots&\vdots\\
0&\cdots&\beta_n&0&\cdots&a_n\alpha_n
\end{pmatrix}\ .
\label{metric_discrete}
\end{align} 

It depends on $2n$ parameters $\alpha_1\ldots\alpha_n$, $\beta_1\ldots\beta_n$. Requirement of positive-definitness of the metric put following conditions on our parameters
\begin{equation}
\alpha_i>0\ ,\quad a_i\alpha_i^2>\beta_i^2\ ,\quad i=1,2,\dots,n\ .
\end{equation}
We can construct corresponding inner product
\begin{equation}
\begin{split}
\langle\langle\psi|\varphi\rangle&=\sum_{i=1}^n{\alpha_i\psi^*_i\varphi_i}\\
&+\sum_{i=1}^n{\beta_i(\psi^*_i\varphi_{n+i}+\psi^*_{n+i}\varphi_i)}\\
&+\sum_{i=1}^n{a_i\alpha_i\psi^*_{n+i}\varphi_{n+i}}\ ,
\end{split}
\end{equation}
where $\psi=(\psi_1,\psi_2,\dots\psi_{2n})^T$, $\varphi=(\varphi_1,\varphi_2\dots,\varphi_{2n})^T$ are complex vectors. 
\section{Conclusions}

In our work, we familiarized the reader with the crypto-Hermitian approach to the Klein-Gordon equation. We computed metric operator in both continuous and discrete cases. Corresponding positive definite inner product for free Klein-Gordon equation was also computed. That is considered a crucial step in proper probability interpretation of the Klein-Gordon equation.  

The next step of this process would be construction of appropriate metric operator for the Klein-Gordon equation with nonzero potential $V$ as was done for special cases in \cite{znojil2004relativistic, znojil2004pseudo, mostafazadeh2006quantum1, mostafazadeh2006quantum2}. It is also possible to broaden the formalism by adding manifest non-Hermiticity in operator $K\neq K^\dagger$, as was shown in \cite{MZdots}.

Related complicated problems with locality, definition of physical observables and attempts to construct conserved four-current can be thoroughly studied in further references \cite{mostafazadeh2006quantum1, mostafazadeh2006quantum2}. The problems become much simpler if we narrow our attention to real Klein-Gordon fields only. It was shown that in such a case, inner product is uniquely defined \cite{mostafazadeh2006quantum1,kleefeld2006some}.

{\bf Acknowledgements:}
The work of Iveta Semor\'adov\'a was supported by the CTU grant Nr. SGS16/239/OHK4/3T/14.

\bibliography{literatura}

\begin{thebibliography}{10}
\providecommand{\url}[1]{\texttt{#1}}
\providecommand{\urlprefix}{URL }
\providecommand{\eprint}[2][]{\url{#2}}

\bibitem{Klein}
O.~Klein.
\newblock Quantentheorie und f{\"u}nfdimensionale relativit{\"a}tstheorie.
\newblock \textit{Zeitschrift f{\"u}r Physik} \textbf{37}(12):895--906, 1926.

\bibitem{Gordon1}
W.~Gordon.
\newblock Der comptoneffekt nach der {Schr{\"o}dingerschen} theorie.
\newblock \textit{Zeitschrift f{\"u}r Physik} \textbf{40}(1-2):117--133, 1926.

\bibitem{Kudar}
J.~Kudar.
\newblock Zur vierdimensionalen formulierung der undulatorischen mechanik.
\newblock \textit{Annalen der Physik} \textbf{386}(22):632--636, 1926.

\bibitem{Fock1}
V.~Fock.
\newblock {\"U}ber die invariante form der wellen-und der bewegungsgleichungen
  f{\"u}r einen geladenen massenpunkt.
\newblock \textit{Zeitschrift f{\"u}r Physik} \textbf{39}(2-3):226--232, 1926.

\bibitem{Fock2}
V.~Fock.
\newblock Zur {Schr{\"o}dingerschen} wellenmechanik.
\newblock \textit{Zeitschrift f{\"u}r Physik A Hadrons and Nuclei}
  \textbf{38}(3):242--250, 1926.

\bibitem{Donder}
T.~De~Donder, et~al.
\newblock La quantification d\'{e}duite de la gravifique {E}insteinienne.
\newblock \textit{Comptes rendus} \textbf{183}:22--24, 1926.

\bibitem{Schrodinger}
E.~Schr{\"o}dinger.
\newblock Quantisierung als eigenwertproblem.
\newblock \textit{Annalen der physik} \textbf{385}(13):437--490, 1926.

\bibitem{constantinescu2013problems}
F.~Constantinescu, et~al.
\newblock \textit{Problems in Quantum Mechanics}.
\newblock Elsevier, 2013.

\bibitem{pauli1934}
W.~Pauli, et~al.
\newblock {\"U}ber die quantisierung der skalaren relativistischen
  wellengleichung.
\newblock \textit{Helv Phys Acta} \textbf{7}:709--731, 1934.

\bibitem{MFKGtype}
A.~Mostafazadeh.
\newblock Hilbert space structures on the solution space of {Klein-Gordon} type
  evolution equations.
\newblock \textit{Class Quantum Grav} \textbf{20}:155--171, 2003.

\bibitem{die}
J.~Dieudonn{\'e}.
\newblock Quasi-hermitian operators.
\newblock \textit{Proc Internat Sympos Linear Spaces (Jerusalem, 1960),
  Pergamon, Oxford} \textbf{115122}, 1961.

\bibitem{dyson}
F.~J. Dyson.
\newblock General theory of spin-wave interactions.
\newblock \textit{Phys Rev} \textbf{102}(5):1217, 1956.

\bibitem{SGH}
F.~Scholtz, et~al.
\newblock Quasi-hermitian operators in quantum mechanics and the variational
  principle.
\newblock \textit{Ann Phys} \textbf{213}:71--101, 1992.

\bibitem{BB}
C.~M. Bender, et~al.
\newblock Real spectra in non-hermitian hamiltonians having
  $\mathcal{PT}$-symmetry.
\newblock \textit{Phys Rev Lett} \textbf{80}(24):5243, 1998.

\bibitem{MFKG}
A.~Mostafazadeh.
\newblock Quantum mechanics of {Klein-Gordon}-type fields and quantum
  cosmology.
\newblock \textit{Ann Phys (New York)} \textbf{309}:1--48, 2004.

\bibitem{mostafazadeh2006quantum1}
A.~Mostafazadeh, et~al.
\newblock Quantum mechanics of {K}lein--{G}ordon fields {I}: Hilbert space,
  localized states, and chiral symmetry.
\newblock \textit{Ann Phys} \textbf{321}(9):2183--2209, 2006.

\bibitem{mostafazadeh2006quantum2}
A.~Mostafazadeh, et~al.
\newblock Quantum mechanics of {K}lein--{G}ordon fields {II}: Relativistic
  coherent states.
\newblock \textit{Ann Phys} \textbf{321}(9):2210--2241, 2006.

\bibitem{mostafazadeh2006physical}
A.~Mostafazadeh.
\newblock A physical realization of the generalized $\mathcal{PT}$-,
  $\mathcal{C}$-, and $\mathcal{CPT}$-symmetries and the position operator for
  {Klein-Gordon} fields.
\newblock \textit{International Journal of Modern Physics A}
  \textbf{21}(12):2553--2572, 2006.

\bibitem{znojil2004relativistic}
M.~Znojil.
\newblock Relativistic supersymmetric quantum mechanics based on {Klein-Gordon}
  equation.
\newblock \textit{J Phys A: Math Gen} \textbf{37}:9557--9571, 2004.

\bibitem{MZdots}
M.~Znojil.
\newblock Solvable relativistic quantum dots with vibrational spectra.
\newblock \textit{Czech J Phys} \textbf{55}:1187--1192., 2005.

\bibitem{znojil2004pseudo}
M.~Znojil, et~al.
\newblock Pseudo-hermitian approach to energy-dependent {Klein-Gordon} models.
\newblock \textit{Czech J Phys} \textbf{54}(10):1143--1148, 2004.

\bibitem{FV}
H.~Feshbach, et~al.
\newblock Elementary relativistic wave mechanics of spin 0 and spin 1/2
  particles.
\newblock \textit{Rev Mod Phys} \textbf{30}(1):24, 1958.

\bibitem{foldy1956synthesis}
L.~L. Foldy.
\newblock Synthesis of covariant particle equations.
\newblock \textit{Phys Rev} \textbf{102}(2):568, 1956.

\bibitem{MFshrnuti}
A.~Mostafazadeh.
\newblock Pseudo-hermitian representation of quantum mechanics.
\newblock \textit{Int J Geom Meth Mod Phys} \textbf{7}:1191--1306, 2010.

\bibitem{bender2007making}
C.~M. Bender.
\newblock Making sense of non-hermitian hamiltonians.
\newblock \textit{Reports on Progress in Physics} \textbf{70}(6):947, 2007.

\bibitem{3HS}
M.~Znojil.
\newblock Three-hilbert-space formulation of quantum mechanics.
\newblock \textit{Symmetry, Integrability and Geometry: Methods and
  Applications} \textbf{5}(001):19, 2009.

\bibitem{kretschmer2004quasi}
R.~Kretschmer, et~al.
\newblock Quasi-hermiticity in infinite-dimensional hilbert spaces.
\newblock \textit{Phys Lett A} \textbf{325}(2):112--117, 2004.

\bibitem{ZnojilKniha}
F.~Bagarello, et~al.
\newblock \textit{Non-selfadjoint operators in quantum physics: Mathematical
  aspects}.
\newblock John Wiley \& Sons, 2015.

\bibitem{znojil2012n}
M.~Znojil.
\newblock N-site-lattice analogues of {$V(x)=ix^3$}.
\newblock \textit{Ann Phys} \textbf{327}(3):893--913, 2012.

\bibitem{kleefeld2006some}
F.~Kleefeld.
\newblock On some meaningful inner product for real {Klein-Gordon} fields with
  positive semi-definite norm.
\newblock \textit{Czech J Phys} \textbf{56}(9):999--1006, 2006.

\end{thebibliography}
\bibliographystyle{abbrv}

\end{document}